\documentclass[twocolumn,showpacs,preprintnumbers,amsmath,amssymb]{revtex4}
\usepackage{graphicx}% Include figure files
\usepackage{dcolumn}% Align table columns on decimal point
\usepackage{bm}% bold math

\begin{document}

\preprint{APS/123-QED}

\title{Core Collapse Via Coarse Dynamic Renormalization}
% Force line breaks with \\

\author{Andras Szell}
%\email{szell@astro.rit.edu}
\author{David Merritt}
%\email{merritt@astro.edu}
\affiliation{Rochester Institute of Technology, 54 Lomb Memorial Drive, 
Rochester, NY 14623}

\author{Ioannis G. Kevrekidis}
\affiliation{Department of Chemical Engineering and Program in Applied and Computational Mathematics, Princeton University,
Princeton, NJ 08544}
%\email{yannis@princeton.edu}

\date{\today}% It is always \today, today,

\begin{abstract}
In the context of the recently developed ``equation-free'' approach
to computer-assisted analysis of complex systems,
we extract the self-similar solution describing core collapse
of a stellar system from numerical experiments.
The technique allows us to side-step the core ``bounce''
that occurs in direct $N$-body simulations due to the small-$N$
correlations that develop in the late stages of collapse,
and hence to follow the evolution well into the self-similar regime.
\end{abstract}

\pacs{Valid PACS appear here}% PACS, the Physics and Astronomy
\maketitle

\section{Introduction}

In many areas of current research,
physical models are available at a fine, 
microscopic scale, while the questions
of most interest concern the system 
behavior on a coarse-grained, macroscopic level.
An example is the gravitational $N$-body problem
\cite{aarseth:03}.
Coarse-grained approximations exist, e. g.
the orbit-averaged Fokker-Planck equation 
\cite{spitzer:87}, which approximates the full 6N
equations of motion by a differential operator
that acts on the single-particle distribution function
$f$.
But the derivation of the Fokker-Planck equation
from the full $N$-body equations of motion requires a
number of approximations.
The ``equation-free'' computational framework
\cite{kevr1,kevr2,kevr3} has recently been
proposed for the computer-assisted study of such
problems, circumventing the derivation of
explicit coarse-grained equations.
The coarse-grained behavior is analyzed directly, 
through appropriately designed short computational
experiments by the fine-scale models.
In the case of problems where the macroscopic behavior
is characterized by scale invariance, dynamic renormalization
\cite{papa:1,papa:2,papa:3},
combined with equation-free computation and a template-based
approach \cite{rowley:05},
can in effect convert the self-similar problem into a 
stationary one,
by working in a frame of reference that expands or
shrinks along with the macroscopic system observables.

Here we apply equation-free dynamic renormalization techniques
to the problem of gravitational core collapse \cite{henon:61,henon:65}.
A star cluster or galaxy evolves due to gravitational
encounters between the stars, which drive the local velocity
distribution toward a Maxwellian.
Stars in the high-velocity tail of the Maxwellian
escape from the system; the probability of escape
in one relaxation time $t_r$,
\begin{equation}
t_r \equiv {0.065 v_m^3\over \rho m G^2\ln\Lambda},
\label{eq:tr}
\end{equation}
is roughly 1\%.
Here $v_m^2$ is the mean square (3D)
velocity of the stars, $\rho$ is the mass density,
$m$ is the individual stellar mass, $G$ is the gravitational
constant, and $\ln\Lambda\approx \ln(0.4N)$
is the Coulomb logarithm \cite{spitzer:87}, with
$N$ the number of stars in the cluster.
Escape occurs primarily from the high-density central
region, or ``core''; if the density contrast
between core and envelope is sufficiently
great, the core exhibits the negative specific heat
characteristic of gravitationally bound systems \cite{wood:68}
and contracts.

Treatments of core collapse based on fluid 
\cite{eggleton:80,louis:91}
 or Fokker-Planck \cite{cohn:79,cohn:80,heggie:88} approximations
to the full $N$-body equations of motion
suggest that the late stages are self-similar,
\begin{subequations}
\begin{eqnarray}
\rho(r,t) &\approx& \rho_c(t)\rho_*\left[{r\over r_c(t)}\right], \\
\rho_c(t) &=& \rho_{c,0}\left(1-t/t_{cc}\right)^\gamma, \\
r_c(t) &=& r_{c,0}\left(1-t/t_{cc}\right)^\delta, \\
\rho_*(r) &\rightarrow& r^{-\alpha},\ \ \ r\gg r_c, 
\end{eqnarray}
\end{subequations}
with $\rho_c$ and $r_c$ the central density and core radius
respectively, and $t_{cc}$ the time at which the central density
diverges; $t_{cc}-t$ is roughly $330$ times the relaxation time
of Eq. 1 if the latter is evaluated at the center of the cluster.

These predictions are in reasonable agreement with the
results of direct $N$-body integrations 
\cite{spurzem:96,makino:96,baumgardt:03}.
But when $N$ is finite, the number of particles in the
core drops to zero as collapse proceeds,
causing two-body and higher order correlations to
dominate the evolution; typically, binary stars form
which halt the collapse (``core bounce'').
In all $N$-body simulations published to date,
this occurs before or only shortly after the core
has entered the self-similar regime.
The rate of binary formation per unit volume due to three-body 
interactions is \cite{hut:85}
\begin{equation}
\dot{n}_{\rm 3-b}\approx 1.2\times 10^2 {G^5m^2\rho^3\over v_m^9}.
\end{equation}
By the time that the number of stars in the core
has dropped to $N_c$ ($N_c\equiv \rho_cr_c^3$), 
the total number of binaries formed is $\sim 10^3 N_c^{-2}$.
Since a single hard binary can halt the collapse,
bounce occurs when $N_c$ has dropped to $\sim 30$.
This limits the maximum central density that 
can be reached in an $N$-body simulation
to a multiple $\sim 10^{3.5}(N/10^4)^3$ of the initial 
density (assuming the initial state described below); 
for $N\approx 10^4$, the achievable density contrast is 
$\sim 10^4$, which is also roughly where
self-similar behavior first appears \cite{cohn:80}. 

In this paper, we exploit coarse dynamic renormalization to 
compute representative self-similar solutions at scales
``away from'' the latest stages of core collapse.
This allows us to avoid the finite-$N$ correlations
that develop at those stages and recover features of
the self-similar behavior predicted at the large-$N$ limit.
We define the ``macroscopic'' quantity of
interest to be the single-particle distribution function
$f(E)$, $E=v^2/2 + \Psi(r)$, with $\Psi(r)$ the self-consistent
gravitational potential (spherical symmetry is assumed throughout).
By dynamically renormalizing the $N$-body model after
each short ``burst'' of integration, we are able to 
maintain a large ($\sim 10^3$) number of particles in the core
even with modest ($\sim 16K$) total numbers,
effectively halting binary formation and
allowing us to accurately extract the form of the 
self-similar solution.

\section{Description of the Calculations}
We adopted the standard Plummer model \cite{plummer:11} initial
conditions for this problem, with density,
gravitational potential, and single-particle distribution function
given by 
\begin{eqnarray}
\rho(r) &=& {3\over 4\pi}{1\over (1+r^2)^{5/2}},\ \ \ \ 
\Psi(r) = -{1\over (1+r^2)^{1/2}}, \nonumber \\
f(E) &=& {24\sqrt{2}\over 7\pi^3} (-E)^{7/2}, \ \ \ \ E = v^2/2+\Psi.
\label{eq:plum}
\end{eqnarray}
Here and below, the gravitational constant $G$ has been set to one.
The $N$-body algorithm was an adaptation of S. J. Aarseth's {\tt NBODY1}
code \cite{aarseth:99} to the GRAPE-6 special-purpose
computer and included a fourth-order
particle advancement scheme with individual, adaptive, block time steps.
We first used this code to carry out a direct integration until
core bounce of a Plummer model with $N=16384$ particles; 
the results (e.g. evolution of the central density, time of
core bounce) agreed well with published studies using different
$N$-body codes \cite{spurzem:96,makino:96,baumgardt:03}.
We used the same number of particles in the calculations described 
below.

Our coarse renormalization procedure consisted of short bursts 
of simulation in a ``lift-simulate-restrict-rescale-truncate'' 
procedure, repeated in a loop until the asymptotic form of
the self-similar solution emerged.
In detail, the steps were:

\noindent
1. {\bf Lift:} Given a smooth representation $\rho(r)$ of the
density profile, a set of Monte-Carlo positions and velocities
was generated, as follows.
First an estimate of the gravitational potential 
$\Psi(r)$ was computed from $\rho(r)$ via Poisson's equation.
The isotropic, single-particle distribution function $f(E)$ 
corresponding to this density-potential pair is given 
uniquely by Eddington's formula \cite{eddington:16}.
For the purposes of generating a new Monte-Carlo set of 
positions and velocities, the relevant function is not $f(E)$ but
$N(<v,r)$, the cumulative number of stars with velocities
less than $v$ at radius $r$.
Using Eddington's formula, this can be shown to be
\begin{eqnarray}
&&N(<v,r) = 1 - {1\over \rho}\int^{E}_0 d\Psi' {d\rho\over d\Psi'} 
\times \nonumber \\
&&\left\{ 1 + {2\over\pi} \left[{v/\sqrt{2}\over\sqrt{E-\Psi'}} - 
\tan^{-1}\left({v/\sqrt{2}\over\sqrt{E-\Psi'}}\right)\right]\right\}.
\end{eqnarray}

\noindent
2. {\bf Integrate.} The $N$-body realization was advanced in 
time until the central density had increased by a factor of 
$\sim 5$ compared with the initial model.
In the nearly-self-similar regime, this required
approximately 250 central relaxation times.

\noindent
3. {\bf Restrict.} An estimate of $\rho(r)$ was 
computed from the particle coordinates ${\bf x}_i, i=1...N$ at the
final time step.
The position of the cluster center was first determined as in
\cite{casertano:85} and the particle radii $r_i$ 
were defined with respect to this center.
Each particle was then replaced by the kernel function
\begin{eqnarray}
K(r,r_i,h_i)&=&{1\over 2(2\pi)^{3/2}}\left({rr_i\over h_i^2}\right)^{-1}
e^{-\left(r_i^2+r^2\right)/2h_i^2} \nonumber \\
&\times&\sinh\left(rr_i/h_i^2\right) 
\label{eq:kernel}
\end{eqnarray}
which is an average over the sphere of radius $r_i$ of the
3D Gaussian kernel of width $h_i$.
The density estimate,
$\rho(r) = \sum_{i=1}^N h_i^{-3} K(r,r_i,h_i)$,
was then computed on a logarithmic radial grid.
The kernel width $h_i$ associated with the $i$th particle
was determined by first constructing
a pilot estimate of the density via a nearest-neighbor scheme, 
then allowing the $h_i$ to vary as the inverse square root of this 
pilot density \cite{silverman:87}.

\noindent
4. {\bf Rescale:} The density estimate was
rescaled,
$\rho(r) \rightarrow A\rho(Br)$,
according to an algorithm that left the core properties
unchanged in the self-similar regime.
The vertical normalization $A$ was fixed by comparing the values
of the density at $r=0$ at the start and end of the integration interval.
The radial scale factor $B$ was adjusted such that the two density
estimates had the same value at the radius where the
density was $1/50$ of its central value.

\noindent
5. {\bf Truncate:} In the late stages of core collapse,
the system develops a $\rho\sim r^{-\alpha}$, $\alpha\approx 2$ 
envelope around the shrinking core.
In direct $N$-body integrations, this envelope grows
to contain most of the total mass, and the number of stars in 
the core drops to zero, leading to the small-$N$ effects
discussed above.
We avoided this by truncating the rescaled density 
beyond a radius $r_1\approx 50 r_c$
using a function that falls smoothly to zero at a radius 
$r_2\approx 10 r_1$.

\noindent
6. Step 1 was then repeated in a loop.

\begin{figure}
\includegraphics[angle=0.,width=7.0cm]{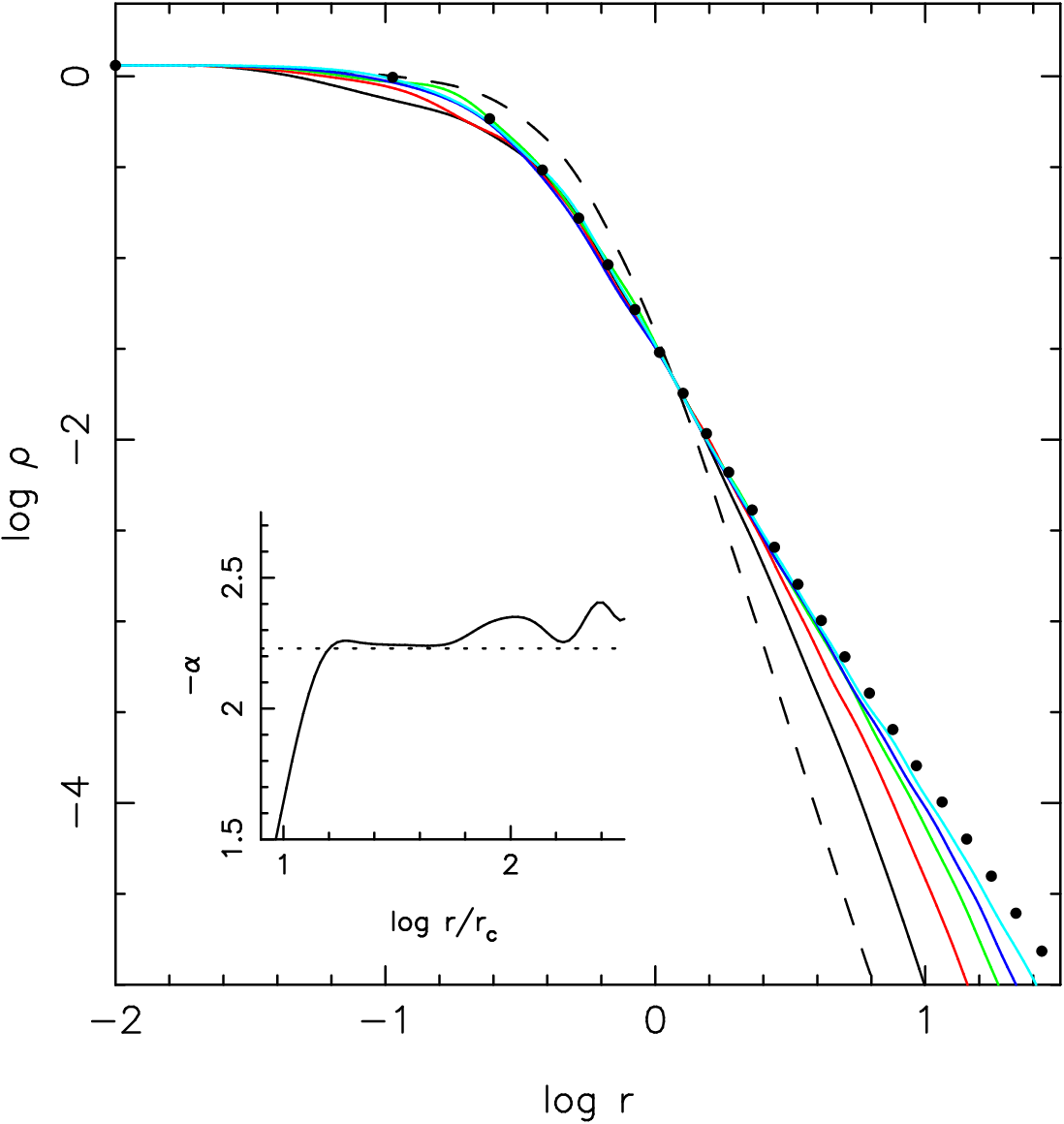}
\caption{\label{fig:f1} Evolution of the rescaled density.
Dashed line is the initial state and points are the 
self-similar $\rho_*(r)$ derived in Ref. \cite{heggie:88}.
Inset shows $\alpha\equiv -d\log\rho/d\log r$ at the
end of the final integration; dashed line is
$\alpha=-2.23$.
}
\end{figure}

Generation of the new particle coordinates from the single-particle
$f(E)$ in step 1 has the effect of removing any binaries that
may have formed in the previous integration interval, and of
resetting to zero any anisotropies that developed in the velocities.
Since the number of particles in the core after each rescaling
was $\sim 10^3$, the chance that even a single
binary would form before the next rescaling was negligible.
Velocity anisotropy was evaluated at the end of each interval
and found to be very small, and restricted to the largest radii.
The effects of binary formation and anisotropy growth could
be reduced still more by increasing $N$ and by reducing the 
lengths of the integration intervals.
Newton-type fixed point algorithms that converge on the
self-similar solution are also possible.

\section{Results}

Fig. 1 shows the density profile at the
end of each rescaling step, 
compared with the self-similar $\rho_*(r)$
computed from the orbit-averaged
isotropic Fokker-Planck equation \cite{heggie:88},
which has $\rho_\star \propto r^{-2.23}$ at large radii.
The renormalized density is quite close to the self-similar
solution after 4-5 iterations.

\begin{figure}
\includegraphics[angle=0.,width=7.0cm]{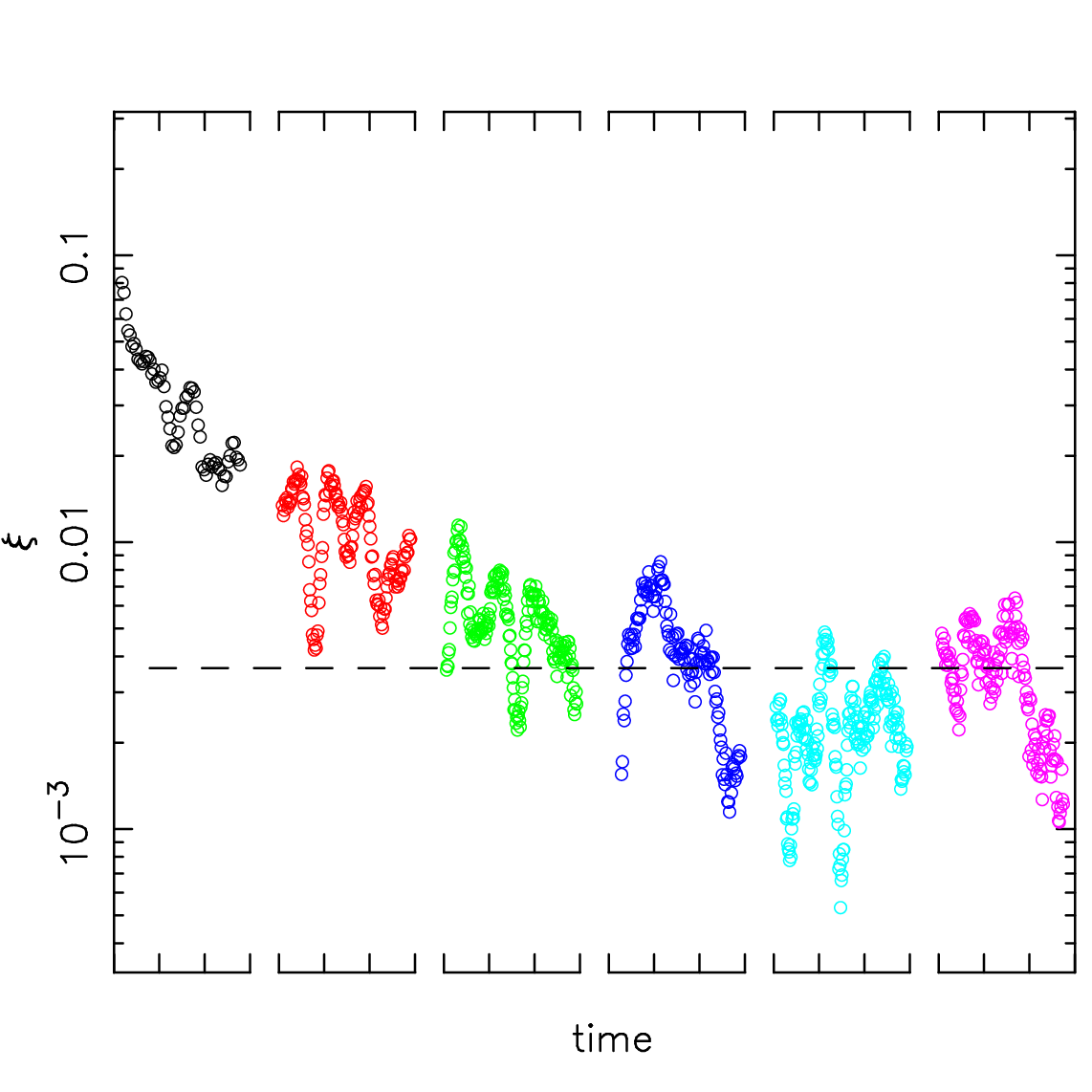}
\caption{\label{fig:f2} 
Evolution of the dimensionless collapse rate parameter 
$\xi\equiv (\dot\rho_c/\rho_c)/t_r$.
Each time interval corresponds to one ``burst'' of integration,
after which the model was rescaled as described in the text.
Dashed line shows the asymptotic (self-similar) value 
$\xi=3.6\times 10^{-3}$
as computed via the isotropic orbit-averaged Fokker-Planck equation 
\cite{cohn:80,heggie:88}.
}
\end{figure}

In the self-similar regime, the finite changes
in central density and core radius during one integration
interval satisfy 
$\log(\rho_{c,f}/\rho_{c,i})/\log(r_{c,f}/r_{c,i})=\gamma/\delta$;
in the Fokker-Planck approximation \cite{heggie:88},
this ratio is equal to the asymptotic density slope,
$\gamma/\delta=\alpha\approx -2.23$.
We computed this ratio from the rescaling factors
($A,B$) in each integration interval and found it to be
$(-2.09,-2.15,-2.23,-2.24,-2.23,-2.23)$, consistent  
with the Fokker-Planck prediction at late times,
and also consistent with our computed value of $\alpha$
(Fig. 1).

The dimensionless collapse rate is 
\begin{equation}
\xi\equiv t_{r}(0){1\over\rho_c}{d\rho_c\over dt}
\end{equation}
with $t_{r}(0)$ the value of the relaxation time (Eq. 1)
evaluated at $r=0$; in the self-similar regime,
$\xi$ should reach a constant value.
Fig. 2 shows the evolution of $\xi$, computed by fitting a 
smoothing spline to $\rho_c(t)$ during each of the integration 
intervals.
Isotropic Fokker-Planck treatments \cite{cohn:80,heggie:88} find
$\xi\approx 3.6\times 10^{-3}$ in the self-similar regime,
consistent within the noise with Fig. 2.

\begin{figure}
\includegraphics[angle=0.,width=7.0cm]{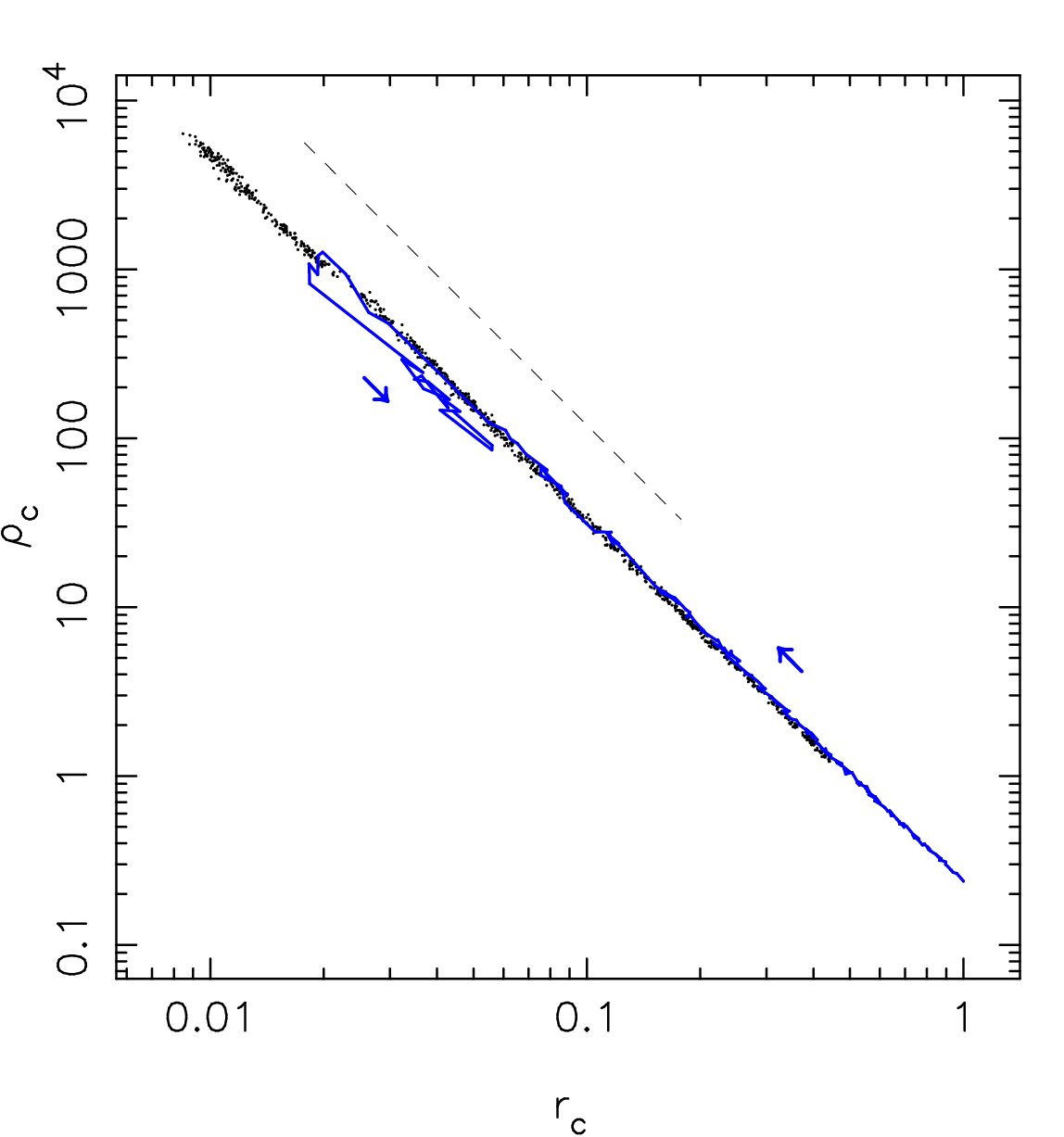}
\caption{\label{fig:f3} 
Central density vs core radius in the rescaled runs
(points) and in a direct $N$-body integration (blue  line).
Dashed line shows $\rho_c\propto r_c^{-2.23}$.
In the direct $N$-body simulation without rescaling,
the central density reaches a peak value then decreases
(``core bounce'') due to formation of binary stars.
}
\end{figure}

Another way to measure the degree of evolution achieved
via the rescaled integrations is shown in Fig. 3.
In this plot, the cumulative effect of the rescalings
on $\rho_c$ and $r_c$ is included; the central
density has increased by a cumulative factor of $\sim 10^{4.5}$ 
by the end of the fifth rescaling,
and the core radius has decreased by a factor $\sim 10^2$.
The $\rho_c(r_c)$ relation from the $N$-body integration
without rescaling is also shown.
In the un-rescaled integration, the central density peaks
due to binary formation at a value $\sim 10$ times lower
than in the rescaled runs.

In principle, one can extract the similarity
exponents $\gamma$ and $\delta$ of Eq. 2 
(not just their ratio)
from the numerical models.
For instance,
if $t_1$ and $t_2$ are two
distinct times in the self-similar regime, 
then
\begin{equation}
\gamma = {t_2-t_1\over {\rho_c(t_2)\over d\rho_c/dt|_{t_2}} - 
{\rho_c(t_1)\over d\rho_c/dt|_{t_1}}}
\end{equation}
and similarly for $\delta$.
Another approach \cite{rowley:05} is via numerical
calculation of the scaling exponents that characterize
the effective differential operator.
A third approach is simply to fit Eqs. 1 to the time-dependent
density and core radius in the self-similar regime.
We had limited success with each of these methods due
to noise associated with the modest particle numbers.
The noise could be reduced by averaging the results from 
different random realizations of the same initial conditions,
or by increasing $N$.

\section{Conclusions}

Coarse dynamic renormalization is a tool for
numerically extracting self-similar solutions by
evolving a dynamical system to a stationary state,
in a scaled reference frame where the self similarity
has been ``factored out.''
We have demonstrated the usefulness of the method
for studying gravitational core collapse.
In the limit of large particle numbers, core collapse
is characterized by a central density that increases 
without limit, due to the combined effects of heat transfer 
and the negative specific heat of the core.
In numerical simulations with finite $N$,
collapse is halted when the number of particles
in the shrinking core drops to a small enough value
that binaries can form.
By carrying out short bursts of integration of 
appropriately rescaled initial conditions, 
we showed that it is possible to
avoid these small-$N$ effects and to follow
collapse well into the self-similar regime.
Furthermore we achieved this with quite modest
particle numbers, $N\approx 10^4$.
In our approach, the degree to which core collapse
can be followed is essentially independent of
the number of particles used, while in direct $N$-body
simulations, the time to core bounce is determined
by $N$.

We chose the macroscopic function of interest 
to be the isotropic, single-particle distribution function 
$f(E,t)$.
The same approximation is commonly made in
Fokker-Planck and fluid treatements of core
collapse 
\cite{eggleton:80,cohn:80,heggie:88}.
A larger particle number would allow us to
relax the assumption of isotropy and extract
$f(E,L,t)$ with $L$ the orbital angular momentum
per unit mass.
Results so obtained could be compared with 
solutions to the anisotropic 
Fokker-Planck \cite{cohn:79,takahashi:95,joshi:00}
and fluid \cite{louis:91} equations.

\begin{acknowledgments}
This work was supported by grants AST-0206031, AST-0420920, 
AST-0437519, and CTS-0205484 from the NSF, 
grants NNG04GJ48G and NAG 5-10842 from NASA, 
and grant HST-AR-09519.01-A from STScI.
\end{acknowledgments}

\bibliography{ms}

\end{document}